\documentclass[a4paper,12pt]{article}

\usepackage{amsmath}
\usepackage{amssymb, amsfonts, amsthm, amscd}
\numberwithin{equation}{section}
\usepackage{bm}
\usepackage{graphicx}
\usepackage{slashed}
\usepackage{xcolor}
\usepackage[all]{xy}
\usepackage{enumitem}
\usepackage{microtype}
\usepackage[hidelinks]{hyperref}
\hypersetup{pdftitle={A Duality Model in Which Dual QCD Becomes Hydrodynamics},
  pdfauthor={Noriaki Aibara, So Katagiri, Yoshiki Matsuoka, Kazuya Mitsutani, Akio Sugamoto}}
\setlist[enumerate]{itemsep=0.5\baselineskip, topsep=0.5\baselineskip,
  parsep=0pt, listparindent=\parindent}
\allowdisplaybreaks[1]
\makeatletter
\def\ps@preprint{\def\@oddhead{\normalfont\shortstack[l]{OCHA-PP-387\\OUJ-FTC-27}}\let\@evenhead\@oddhead\let\@oddfoot\@empty\let\@evenfoot\@empty}
\makeatother
\begin{document}

\thispagestyle{preprint}
\begin{center}
{\Large\bfseries
A Duality Model in Which Dual QCD Becomes Hydrodynamics}

\baselineskip 18pt
\vspace{7mm}

{\bfseries Noriaki Aibara$^{1}$, So Katagiri$^{2}$, Yoshiki Matsuoka$^{1}$,\\
Kazuya Mitsutani$^{1}$, Akio Sugamoto$^{3}$}

\vspace{4mm}
{\small\itshape
\setlength{\baselineskip}{13pt}
$^{1}$Nature and Environment, Faculty of Liberal Arts,\\
The Open University of Japan, Chiba 261-8586, Japan\\
$^{2}$Major in Complex Systems Science, Graduate School of Science and Engineering,\\
Ibaraki University, Nakanarusawa-cho, Hitachi, 316-8511, Japan\\
$^{3}$Department of Physics, Graduate School of Humanities and Sciences, Ochanomizu University,\\
2-1-1 Otsuka, Bunkyo-ku, Tokyo 112-8610, Japan
}
\end{center}

\vspace{3mm}
\begin{abstract}
\normalfont
Revisiting a duality model in which dual QCD becomes hydrodynamics, or the dual gluon is a fluid, allows us to understand some of the mysteries of QCD, such as the behavior of the quark--gluon plasma (QGP) as a perfect fluid and gluon saturation in the structure function at high energies. This old-fashioned duality converts the gluon action into hydrodynamics described by the 2-form stream function $W^a_{\mu\nu}$ with eight colors. All interactions then become current--current interactions of the Nambu--Jona-Lasinio type. The resulting theory admits a strong-coupling expansion about the fluid. This can provide a third method to supplement perturbative QCD and lattice QCD.

\end{abstract}

\section{Introduction}
QCD is a fundamental theory of strong interactions. Its development began in 1965, when Y. Nambu formulated an $SU(3)$ gauge theory for quarks with his ``charm'' quantum number (now called ``color'') and showed, by calculating Casimir invariants, that baryons and mesons are stable when they are color singlets~\cite{NambuQCD}. After the discovery of asymptotic freedom in 1973~\cite{asymptotic free}, analytical studies of QCD progressed rapidly. Since the late 1970s, numerical calculations based on lattice gauge theory have drastically changed hadron physics. Lattice gauge theory was introduced by
K. Wilson in 1974~\cite{lattice gauge theory}. In lattice gauge theory, quark confinement and a linearly rising potential between a quark and an antiquark are derived extremely simply at the lowest nontrivial order of the strong-coupling expansion in $g_s$. Ironically, this simplicity makes it difficult to grasp the physical reasoning and picture behind the lattice result.

This gap was filled by 't Hooft in 1978~\cite{t Hooft}, who proposed an order--disorder theory of QCD. He introduced a disorder parameter $B(C')$ defined on a curve $C'$, in addition to the order parameter given by the Wilson operator $A(C)$ on a curve $C$, and discussed the well-known area and perimeter laws for $A$ and $B$. This order--disorder theory clarified that the ordered phase, or Higgs phase (with massive quarks), arises through the condensation of electric objects, whereas the disordered phase, or confined phase, arises through the condensation of magnetic objects. This marks the beginning of the concept of duality: exchanging electricity and magnetism. However, it has not yet been established which magnetic objects (magnetic monopoles, vortices, instantons, or others) are responsible for confinement.

At present, numerical and analytical approaches are progressing and stimulating each other, although many unresolved issues remain. See, for example, Refs.~\cite{QCD1,QCD2,CERN}.

In this context, it is useful to consider a third approach, in addition to perturbative QCD and lattice QCD, that provides an analytical treatment through a strong-coupling expansion in QCD. This approach was proposed by one of the authors (A.S.) in 1978. The motivation was to apply order--disorder theory in solid-state physics to four-dimensional continuum gauge theories, noting the similarity between order--disorder theory and quantum mechanics. The uncertainty relation between coordinates and momenta $(x, p)$ is a form of order--disorder relation. A Fourier transformation is therefore applied to exchange the electric and magnetic representations. The resulting theory is described in terms of 2-form fields. Although this description initially seemed unfamiliar, its interpretation became clear after Nambu's talk in Tokyo in 1978. The theory is hydrodynamics based on Kalb--Ramond 2-form gauge theory.

We must also recall that the lattice gauge theory action is \emph{periodic}, unlike the usual continuum gauge theory action. Schematically, the lattice action is
\begin{align}
S_{\text{lattice}} \propto \sum_x \sum_{\mu\nu} \operatorname{Tr} (e^{i\bm{F}_{\mu\nu}(x)} -1),
\end{align}
where
$\bm{F}_{\mu\nu}(x)$, defined by $\bm{F}_{\mu\nu}(x) =g_s a^2 (T^a F^a_{\mu\nu}(x))$, is the flux through an elementary square plaquette in the $(\mu\nu)$ directions at $x$. As usual, $g_s$ is the strong coupling constant, $a$ is the lattice spacing, and $T^a$ and $F^a_{\mu\nu}(x)$ are the $SU(3)$ generators and field strength, respectively.

The lattice action is periodic under shifts of the flux by $2\pi$, with the symmetry $\bm{F}_{\mu\nu}(x) \to \bm{F}_{\mu\nu}(x) + 2\pi n_{\mu\nu}(x)$ ($n_{\mu\nu}(x) \in \mathbb{Z}$). In other words, $n_{\mu\nu}(x)$ is a topological number that quantizes the magnetic flux. Thus, lattice gauge theory is invariant under the creation and annihilation of quantized magnetic fluxes with a minimum length $a$. This forms a disordered medium. No such medium appears in continuum QCD.

In this paper, to follow lattice QCD, we introduce periodicity by including a field theory of vortex strings based on the work of
Kalb and Ramond, which describes the creation and annihilation of vortex filaments in dual hydrodynamics.
The periodicity studied here is a special case in which the topological number $n$ is constant along a thin closed curve $C$, or a closed vortex filament.

\section{Application of old-fashioned duality to QCD}

Here, we recall the old-fashioned duality model studied in Refs.~\cite{AS-1,AS-2,AS-3}. This model is obtained by applying a Fourier transformation to a gauge theory:
\begin{align}
&e^{ i \int d^4x \left\{- \frac{1}{4} \varepsilon  F^a_{\mu\nu}(x)^2 \right\} } \nonumber \\
&\propto \int \mathcal{D} W^a_{\mu\nu}(x)
 \; e^{  i \int d^4x \left\{ -\frac{1}{4\varepsilon} W^a_{\mu\nu}(x) W^{a\mu\nu}(x) +\frac{1}{2} \tilde{W}^a_{\mu\nu}(x) F^{a\mu\nu}(x) \right\} }, \label{eq:duality}
\end{align}
Here, $\varepsilon$ denotes the dielectric constant. This transforms the gauge theory into hydrodynamics, where $W^a_{\mu\nu}$ is the stream function in the adjoint representation, its dual tensor is $\tilde{W}^a_{\mu\nu}=\frac{1}{2}\epsilon_{\mu\nu\lambda\rho} W^{a, \lambda\rho}$, and the fluid velocity is $V^a_{\mu}(x)=\partial^{\nu} \tilde{W}^a_{\nu\mu}(x)$, satisfying the continuity condition $\partial_{\mu} V^{a\mu}(x)=0$. Thus, the fluid is incompressible and inviscid. When this transformation is applied to QCD, with gauge group $SU(3)$ and strong coupling constant $g_s$, the dual gluon becomes a perfect fluid.
The hydrodynamic interpretation of $W^a_{\mu\nu}$, a Kalb--Ramond field and a string gauge field~\cite{K-R}, is due to Nambu~\cite{Nambu}. This hydrodynamics has been used in superfluidity~\cite{Hatsuda} and soliton theory~\cite{Inami}.

As mentioned in the introduction, periodicity or magnetic objects are needed to discuss confinement. Here, we choose the magnetic vortex string, or the vortex filament in dual hydrodynamics. To introduce vortex filaments, we add two adjoint scalars, $\phi^a$ and $\psi^a$. In a gauge-fixed description, their vacuum expectation values $\langle\phi^a\rangle_{\mathrm{gf}}$ and $\langle\psi^a\rangle_{\mathrm{gf}}$ serve as order parameters characterizing the scalar background relevant to vortex formation.

For a thin vortex filament, the energy is stored
solely along a thin curve $C$, so it can be represented as a string with tension $M^2$ (energy per unit length). It sweeps out a world sheet $X^{\mu}=X^{\mu}(\tau, \sigma)\; (-\infty < \tau < \infty, 0 \le \sigma < 2\pi)$. We consider a closed string of arbitrary shape. The probability amplitude associated with a given vortex-filament configuration $X(\tau,\sigma)$ is described by the string wave functional $\Psi[X(\tau,\sigma)]$.

As mentioned above, the scalars act as \emph{stirring rods} that create vortices.\footnote{In the high-energy limit, $Q^2\to\infty$ and $g_s(Q^2)\to0$, the dual gluon is described as a perfect fluid. At lower energies, non-Abelian interactions are expected to induce shear stress. Here, shear stress denotes anisotropic spatial stress in the local rest frame of the fluid. In this picture, the octet scalars act as stirring rods and generate vortex filaments through this stress.

For the classical current--current term of Eq.~\eqref{eq:dual-qcd}, let
\[
J^{a\mu}=g_s^{-1}V^{a\mu}+j_q^{a\mu}+j_{\phi,\psi}^{a\mu},
\qquad \Pi^a_\mu=M^{ab}_{\mu\nu}J^{b\nu}.
\]
Using signature $(-,+,+,+)$ and varying the metric at fixed $W^a_{\mu\nu}$, scalar fields, and spinor components gives
\[
\begin{aligned}
T^{\mu\nu}_{\mathrm{main}}
={}&h^{ab}\Pi^{a\mu}\Pi^{b\nu}
 -2\Pi^{a(\mu}j_{\phi,\psi}^{a\nu)}
 -\Pi^{a(\mu}j_q^{a\nu)}\\
&+g^{\mu\nu}\left[
\Pi^a_\lambda\bigl(j_{\phi,\psi}^{a\lambda}+j_q^{a\lambda}\bigr)
-\frac12h^{ab}\Pi^a_\lambda\Pi^{b\lambda}\right],
\end{aligned}
\]
where $T^{\mu\nu}_{\mathrm{main}}=2(-g)^{-1/2}\delta S_{\mathrm{main}}/\delta g_{\mu\nu}$ and parentheses denote symmetrization with weight $1/2$. The variation includes the metric dependence of the Hodge dual and of $\gamma^\mu$. For negligible matter currents, this contribution has the perfect-fluid form if the color components share a common timelike velocity. The remaining terms in Eq.~\eqref{eq:dual-qcd} contribute separately to the total energy--momentum tensor. When the inverse admits a strong-coupling expansion, its first-order term is
\[
\delta M^{ab}_{\mu\nu}=-g_s^{-1}(h^{-1})^{ac}f^{cde}\widetilde W^e_{\mu\nu}(h^{-1})^{db},
\]
without assuming that $h^{-1}$ commutes with the color matrix $f\widetilde W$.} However, the main reason for introducing adjoint scalars is to introduce periods into the system.\footnote{A period is a real number with a degree of freedom corresponding to color in QCD, so a scalar in the adjoint representation must be introduced.} These periods are related to the phase periodicity of the scalar fields around vortex filaments. Another role is to regularize the singularity at the vortex center.

After introducing the two scalars, we integrate the gluon fields $A^a_{\mu}(x)$ out of the action.\footnote{More precisely, we perform the dual transformation after fixing the gauge by $\phi^8=\psi^8=\phi^3=\psi^3=0$. We define the complex scalar components by
\[
\begin{aligned}
\Phi_1&=\phi^4+i\phi^5, &\qquad \Phi_2&=\phi^6+i\phi^7,\\
\Psi_1&=\psi^4+i\psi^5, &\qquad \Psi_2&=\psi^6+i\psi^7.
\end{aligned}
\]
The phase winding numbers of $\Phi_1,\Phi_2,\Psi_1,\Psi_2$ are the integers $n_1,n_2,n_3,n_4$, respectively. On circling a vortex, their phases change by $2\pi n_1,2\pi n_2,2\pi n_3,2\pi n_4$, respectively. Mathematically, a vortex filament solution is guaranteed by $\pi_1(SU(3)/\mathbb{Z}_3)=\mathbb{Z}_3$. The quotient by $\mathbb{Z}_3$ arises because the scalars belong to a real representation.} We then obtain the following dual QCD Lagrangian, $\mathcal{L}^{\star}_{\mathrm{QCD}}$:
\begin{align}
&\mathcal{L}^{\star}_{\mathrm{QCD}}
= \frac{1}{2} \left( g_s^{-1}V^{a,\mu} +  j_q^{a\mu} +  j_{\phi,\psi}^{a\mu}  \right) M^{ab}_{\mu\nu} \left(g_s^{-1}V^{b,\nu} +  j_q^{b\nu} +  j_{\phi,\psi}^{b\nu}  \right) \nonumber \\
&- \frac{1}{4} (W^{a}_{\mu\nu})^2 +  \frac{1}{2}\operatorname{Tr}\ln M^{ab}_{\mu\nu}(x) \nonumber \\
&+ ~\mathcal{L}(n_F \; \text{free quarks}) +  \mathcal{L}(\text{two octet scalars}).
\label{eq:dual-qcd}
\end{align}

The fluid velocity $V^a_{\mu}$ is added, with strength $g_s^{-1}$, to the particle currents. The quark current includes a sum over the flavor index:
\[
j_q^{a\mu}=\sum_{i=1}^{n_F}j_{q_i}^{a\mu},
\qquad j_{q_i}^{a\mu}=\bar q_i(x)\gamma^\mu T^a q_i(x).
\]
The scalar currents and their sum are
\[
\begin{aligned}
j_\phi^{a\mu}&=f^{abc}\phi^b\partial^\mu\phi^c,
&\qquad j_\psi^{a\mu}&=f^{abc}\psi^b\partial^\mu\psi^c,\\
j_{\phi,\psi}^{a\mu}&=j_\phi^{a\mu}+j_\psi^{a\mu}.
\end{aligned}
\]
A ``classical mass'' of the fluid is given by
$M^{ab}_{\mu\nu}=\left(h^{ab}g_{\mu\nu} + g_s^{-1} f^{abc} \tilde{W}^c_{\mu\nu}\right)^{-1}$ with $h^{ab}= f^{acd}f^{bce}(\phi^d \phi^e + \psi^d \psi^e)$.

Without the term $-\frac{1}{4} (W^a_{\mu\nu})^2$, free hydrodynamics is invariant under
\begin{align}
&W^{a}_{\mu\nu} \rightarrow W^{a}_{\mu\nu} + \nabla^{ab}_{\mu}\Lambda^{b}_{\nu} - \nabla^{ab}_{\nu}\Lambda^{b}_{\mu},
\enspace \nabla^{ab}_{\mu} = \delta^{ab} \partial_{\mu} + g_s^{-1} f^{acb} P^{c}_{\mu},  \nonumber
\end{align}
The field $P^a_{\mu}= M^{ab}_{\mu\nu} V^{b,\nu}$ enters the connection defining the covariant derivative $\nabla^{ab}_{\mu}$. This construction provides a fluid-dynamical interpretation of the covariant derivative. The non-Abelian formulation is related to the two-form gauge theory discussed by Y. Nambu~\cite{Nambu} and to the non-Abelian antisymmetric-tensor theory developed by D. Z. Freedman and P. K. Townsend~\cite{F-T}.

Under the Kalb--Ramond transformation, the string field of a vortex filament, $\Psi_C[X]$, transforms as follows:
\begin{align}
\Psi_C[X] \to e^{i QT^a \oint dx^{\mu} \Lambda^a_{\mu}(x)}\Psi_C[X].
\end{align}
Since the vortex filament moves with the fluid, Stokes' theorem for a curve deformation $\delta C$ should be expressed using the covariant derivative defined above. The following string Lagrangian of Kalb--Ramond type is therefore invariant:
\begin{align}
&\mathcal{L}^{\star}_{\text{KR}}= \left| \oint_C dx^t \left(\frac{\delta}{\delta X^{\mu t}(x)} - i  QT^a g_s^{-1} W^{a}_{\mu t} (x) \right) \Psi_C[X] \right|^2,
\end{align}
where the deformation $\delta C^{\mu t}$ is a keyboard deformation of a curve in its normal direction $\mu$ and tangential direction $t$.\footnote{
The keyboard deformation refers to an
infinitesimal deformation of a curve that
sweeps out an oriented area element spanned
by the tangential direction $t$ and the
deformation direction $\mu$.
} Here, $Q$ denotes the vortex-string charge and should be distinguished from the momentum scale $Q^2$ used below.

The equation of motion (EOM) for the fluid has an elegant form and admits various solutions:
\begin{align}
\partial_{\mu} P^a_{\nu} - \partial_{\nu} P^a_{\mu} + g_s^{-1} f^{abc} P^b_\mu P^c_\nu =g_s^2 \tilde{W}^a_{\mu\nu}.\label{eq:fluid-eom}
\end{align}
A simple way to find a thin vortex solution is to assume that only the eighth-color components $(V^8_{\mu}, W^8_{\mu\nu})$ are nonzero. The EOM then becomes Abelian-like, and a thin vortex filament solution can be found in the same way as the Nielsen--Olesen solution.
With these Abelian-like vortex filaments, the non-Abelian Kalb--Ramond symmetry is broken down to the $\mathbb{Z}_3$ symmetry of the center of $SU(3)$, generated by $\omega=e^{i \frac{4\pi}{\sqrt{3}} T^8}$.

\section{Characteristics of dual QCD}

The characteristics are summarized as follows:
\begin{enumerate}[label=(C\arabic*), ref=C\arabic*, leftmargin=2.6em]

\item\label{item:C1} To follow lattice QCD, periodicity of the action is introduced into continuum QCD by including closed vortex filaments of various shapes and topological numbers.

The two additional octet scalars determine the periods or topological numbers.\footnote{Color-singlet bound states of the additional color-octet scalars have not been experimentally established. Constraints on their masses depend on the production and decay channels and on the details of the model.}

\item Dual QCD is hydrodynamics, in which the dual gluon is a fluid. There is no viscosity at the classical level.
\item All interactions are Nambu--Jona-Lasinio (NJL)-type~\cite{NJL} current--current interactions with a common weight $M^
{ab}_{\mu\nu}$:
\begin{description}[style=unboxed, leftmargin=0pt, labelsep=0.5em, font=\normalfont]
\item[$M(j_q j_q)$:] the usual quark current--current interaction;
\item[$M(g_s^{-1} V j_q)$:] the fluid-current--quark-current interaction;
\item[$M(j_{\phi,\psi} j_{\phi,\psi})$:] the scalar current--current interaction;
\item[$M(g_s^{-1} V j_{\phi,\psi})$:] the fluid-current--scalar-current interaction.
\end{description}

The scalar current acts as a stirring rod that generates vortex filaments and thus induces a string field for them. Accordingly, $M(g_s^{-1} V j_{\phi,\psi})$ induces the Kalb--Ramond coupling of vortex strings~\cite{AS-3}, while $M(j_{\phi,\psi} j_{\phi,\psi})$ induces a four-string interaction. The following field-theoretical Lagrangian for vortex filaments is then derived:
 \begin{align}
&\mathcal{L}^{\star}_{\text{string, vf}}=\sum_C \left| \oint_C dx^t \left(\frac{\delta}{\delta X^{\mu t}(x)} - i g_s^{-1}QW^8_{\mu t}(x)\right)\Psi_C[X]\right|^2 \nonumber \\
&\quad-\sum_C M^2|\Psi_C[X]|^2-\sum_{C,C'}G_4\left(|\Psi_C[X]|^2|\Psi_{C'}[X']|^2\right).\label{eq:string-vf}
\end{align}
Here, the sum over $C$ symbolically denotes a sum or functional integration over closed vortex-filament configurations.
\item Symmetry of dual QCD: the $SU(3)$ non-Abelian Kalb--Ramond symmetry is broken by both the ``mass term'' $-\frac{1}{4} (W^a_{\mu\nu})^2$ and the Abelian-like vortex configurations, down to Abelian Kalb--Ramond symmetry
with global center symmetry $\mathbb{Z}_3$. The $SU(3)$ gauge symmetry is broken only by the latter, down to the global center symmetry.

\item In the high-energy limit, $Q^2 \to \infty$, the dual gluon dominates over quarks and behaves as an incompressible perfect fluid without viscosity at the classical level. The effective Lagrangian in this limit is
\begin{align}
&\mathcal{L}^{\star}(Q^2\to\infty) \nonumber \\
&\quad=\frac{1}{2g_s^2}\Bigl\{\bigl[h^{de}g^{\alpha\beta}+g_s^{-1}(Q^2)f^{ade}\widetilde W^{a\alpha\beta}\bigr]^{-1}\Bigr\}_{bc,\mu\nu}V^{b\mu}V^{c\nu},\label{eq:high-fluid} \\
&\mathcal{L}^{\star}_{\text{vf}}(Q^2 \to \infty) = \sum_C\left| \oint_C dx^t \left(\frac{\delta}{\delta X^{\mu t}(x)} - i g_s^{-1}QW^8_{\mu t}(x)\right)\Psi_C[X]\right|^2.\label{eq:high-energy}
\end{align}

In Eq.~\eqref{eq:high-fluid}, the outer indices on the braces label the components after inversion in the combined color and Lorentz spaces. The high-energy limit $g_s(Q^2)\to0$ should be distinguished from the strong-coupling expansion $g_s^{-1}(Q^2)\ll1$ used below.

\item In the low-energy limit, $Q^2 \to 0$, the effective Lagrangian is the sum of a quark contribution with four-Fermi interactions and a background-vortex contribution with four-string interactions:\footnote{
Here, we assume the strong-coupling regime
$g_s^{-1}(Q^2)\ll 1$ and that $h^{ab}$ is
invertible. Under these assumptions, and
provided that the $g_s^{-1}\widetilde W$
contribution is perturbatively small,
$M$ approaches the inverse of $h g$ at
leading order. The low-energy limit
$Q^2\to 0$ does not by itself imply
$g_s^{-1}\to 0$.
The matrix $h^{ab}$ is invertible for generic
configurations of the two adjoint scalar fields.
More precisely, invertibility requires that the
two adjoint fields have no common nonzero
infinitesimal stabilizer, i.e.,
\[
\ker(\mathrm{ad}_\phi)
\cap\ker(\mathrm{ad}_\psi)=\{0\}.
\]
}
\begin{align}
&\mathcal{L}^{\star}(Q^2\to0)=\sum_{i=1}^{n_F}\bar q_i(i\slashed{\partial})q_i
+\frac12(h^{-1})_{ab}\sum_{i,j}j_{q_i}^{a\mu}j^b_{q_j\mu}
-\frac14(W^a_{\mu\nu})^2,\quad\text{and}\label{eq:low-energy}\\
&\mathcal{L}^{\star}_{\text{vf}}(Q^2\to0)=\sum_C\left|\oint_C dx^t\left(\frac{\delta}{\delta X^{\mu t}(x)}\right)\Psi_C[X]\right|^2
-\sum_C M^2|\Psi_C[X]|^2\nonumber \\*
&\qquad-G_4\sum_{C,C'}\left(|\Psi_C[X]|^2\right)\left(|\Psi_{C'}[X']|^2\right).\label{eq:low-vf}
\end{align}

\item Stueckelberg trick:
Dual QCD breaks Kalb--Ramond (KR) symmetry through the
``mass term'' $-\frac{1}{4} (W^a_{\mu\nu})^2$. The Stueckelberg trick therefore works: a suitable KR transformation can change the form of the interaction.
For example,
\begin{enumerate}[label=\arabic*), leftmargin=2em]
\item Choosing the Landau-gauge KR propagator eliminates the KR coupling and generates a calculable interaction $(W^8_{\mu\nu}(x))^2 \sum_{C \ni x} |\Psi[C]|^2$ (Seo and A.S. in Ref.~\cite{AS-1}).
\item $M (V^a \cdot j_q) \; \Rightarrow M^2 \partial^2 (j_q)^2$.
\par
The current--fluid interaction may be recast into derivative current--current interactions, schematically represented by the relation above. The precise tensor contractions and coefficients are not specified here.
\end{enumerate}

\item Periodicity of dual QCD: an addendum to (\ref{item:C1}).

We apply the inverse dual transformation to the following dual Lagrangian with fixed topological numbers:
\begin{align}
&\mathcal{L}^{\star}_{\text{periodic}}=\frac{1}{2} V^a_{\mu} M^{ab, \mu\nu}  j^b_{\nu} + \cdots \\
&\propto -\tilde{W}^a_{\mu\nu} M(a^{-1})^{\mu} (a^{-1})^{\nu}  \sum_{(b,c)} 2\pi n_{bc}\; |\phi_{bc}|^2 f^{abc} +\cdots,\label{eq:periodic-dual-approx}
\end{align}
where $n_{ab}$ is a topological number specifying how the phase of $\phi^a + i \phi^b$ winds around the vortex filament, and $|\phi_{ab}|^2=(\phi^a)^2+(\phi^b)^2$.\footnote{Here, $(a^{-1})^{\mu}=\delta(x^{\mu})$ in the continuum theory, and the $(\lambda\rho)$ and $(\mu\nu)$ plaquettes are related by duality. Schematically, we have
\[
(a^{-1})^{\lambda}(a^{-1})^{\rho}
=\delta(x-X)^{\lambda}\delta(x-X)^{\rho}
\propto \epsilon_{\mu\nu}{}^{\lambda\rho}
\int d\tau\,d\sigma\,
\frac{\partial(X^{\mu},X^{\nu})}{\partial(\tau,\sigma)}
\delta^{(4)}(x-X).
\]
Here, a current flows around the boundary of the $(\lambda\rho)$ plaquette, generating a magnetic vortex filament whose world sheet is represented by the dual $(\mu\nu)$ plaquette. The Kalb--Ramond field $W^a_{\mu\nu}$ couples to this world sheet. The numerical coefficient depends on the normalization of the plaquette and world-sheet current. Accordingly, Eq.~\eqref{eq:periodic-dual-approx} specifies the displayed contribution only up to an overall factor.} We have used the approximation $M^{ab, \mu\nu} \approx g_s^{-2} (\Phi)^{-1} \delta^{ab} g^{\mu\nu}$, valid at low energies or in the Abelian case.

We then arrive at periodic QCD:\footnote{
Here, the factors
$(a^{-1})^\lambda(a^{-1})^\rho$
are understood as oriented elements
associated with the dual plaquette
in the $(\lambda,\rho)$ directions,
rather than as the ordinary symmetric
product of two components of a single vector.
The numerical coefficient in the vortex contribution to the periodic Lagrangian is schematic and is not fixed by Eq.~\eqref{eq:periodic-dual-approx}.
}
\begin{align}
&\mathcal{L}_{\text{periodic}} =  -\frac{1}{4} \left[ F^a_{\mu\nu}
+\frac12\epsilon_{\mu\nu\lambda\rho}M(a^{-1})^{\lambda}(a^{-1})^{\rho}
\sum_{(b,c)}2\pi n_{bc}\;f^{abc}|\phi_{bc}|^2 \right]^2 \nonumber \\
 &~~~~~~~~~~+ \cdots.
\end{align}

The topological numbers are summed over in the partition functions, yielding
\begin{align}
&\sum_{n_{bc} \in \mathbb{Z}} \int \mathcal{D} A^a_{\mu}(x) \cdots e^{i S_{\text{periodic QCD}}} \nonumber \\
&\propto \sum_{n_{bc} \in \mathbb{Z}} \int \mathcal{D} W^a_{\mu\nu}(x) \cdots e^{i S^{\star}_{\text{dual QCD}}}.
\end{align}
This shows that our dual QCD partly resembles lattice QCD.
\end{enumerate}

\section{Possible implications of the dual QCD model}

We summarize the possible implications of the dual QCD model.
\begin{enumerate}[label=(I\arabic*), ref=I\arabic*, leftmargin=2.6em]

\item In our dual QCD, all interactions are NJL-type $(\text{current})^2$ interactions. This is advantageous for demonstrating the mass gap using the original self-consistent approach, the Coleman--Weinberg method, or the more recent functional renormalization group (FRG) method. It is important to distinguish the roles of the NJL model: in ordinary QCD it is a plausible assumption, whereas in dual QCD it is derived.
Even in dual QCD, interactions between the fluid and fermionic currents must be taken into account carefully. This can, however, be done analytically through a strong-coupling expansion in $g_s^{-1}$.

This theory might suggest a way to make four-Fermi interactions renormalizable through the active contribution of fluid dynamics.

\item The main term in the low-energy effective action given in Eq.~\eqref{eq:low-energy} is
\begin{align}
&\mathcal{L}^{\star}(Q^2\to0)_0=\sum_{i=1}^{n_F}\bar q_i(i\slashed{\partial})q_i\nonumber\\
&\qquad+\frac12(h^{-1})_{ab}\sum_{i,j}(\bar q_i T^a\gamma_\mu q_i)(\bar q_j T^b\gamma^\mu q_j),\label{eq:low-quarks}
\end{align}
which is modified by vortices through the fluid in the strong-coupling expansion in $g_s^{-1} \ll 1$. Vortex condensation must be incorporated when confinement effects become important.

This action can be used to estimate quark and hadron masses.

\item The dual transformation turns the dual gluon into a fluid without viscosity at the classical level. At high energies, fermion currents decouple, leaving a fluid of dual gluons. More precisely, within the strong-coupling regime $g_s^{-1}\ll1$, subdominant $g_s^{-1}$ corrections or quantum effects may generate nonzero viscosity.
This leading-order result may provide a qualitative explanation for the fluid-like behavior of QGP observed in heavy-ion experiments at RHIC and LHC-ALICE.

The vortex-filament action in Eq.~\eqref{eq:high-energy} suggests that vortex rings are pair-produced from the dual gluon fluid through Kalb--Ramond interactions. If so, this pair production is expected to be observable experimentally.

\item Implications for the QCD phase diagram are our next target (work in preparation by N. Aibara \textit{et al.}). The aim is to obtain a phase diagram in a numerically reliable way and compare the result with lattice gauge theories and other approaches. Here, we briefly outline the strategy.

One characteristic of our dual QCD is that all interactions between quarks, as well as those between vortex strings, are of NJL type with a common weight. Various condensates therefore enter in similar forms. For example, the contributions associated with chiral condensation and adjoint-scalar condensation take the following forms, respectively:
\begin{align}
G_4 \langle \bar{q}q \rangle \bar{q}(x) q(x),~g_s \langle \phi^{\dagger} \phi \rangle (A_{\mu}(x))^2.
\end{align}
For string condensation, we have similar relations:
In a mean-field treatment of the four-string interaction, the induced quadratic term can be expressed using an effective coupling $G'_4$, whose relation to $G_4$ depends on the normalization and combinatorial factors in the sum over string configurations.
\begin{align}
&G'_4\langle \Psi^{\dagger} \Psi \rangle \Psi_C[X]^{\dagger} \Psi_C[X]= M^2 \Psi_C[X]^{\dagger} \Psi_C[X].  \\
&-\frac{1}{4} \left[ 1+ g_s^{-2} \langle \Psi^{\dagger} \Psi \rangle \right] (W_{\mu\nu})^2=-\frac{1}{4	\varepsilon} (W_{\mu\nu})^2.
\end{align}
Here, $M^2$ is the vortex-string tension, and $\varepsilon$ is the dielectric constant, which becomes less than one due to vortex condensation:
\begin{align}
\varepsilon =\frac{1}{ 1+ g_s^{-2} \langle \Psi^{\dagger} \Psi \rangle }.
\end{align}
The energy dependence of the condensate determines the running of the coupling $g_s(Q^2)$. See the special case discussed in the 1981 paper in Ref.~\cite{AS-1}.

An effective action can now be obtained by assuming constant chiral and vortex condensates, $\langle \sigma \rangle$ and $\langle \Sigma \rangle$, defined by $\langle \sigma\rangle = \sum_{i=1}^{n_F} \langle \bar{q}_i q_i \rangle, \; \langle \Sigma \rangle= \sum_{C \ni x} \langle \Psi[C]^{\dagger} \Psi[C] \rangle$:
\begin{align}
 S_{\text{eff}}=  \pm \operatorname{Tr} \ln \left[ p_E^2 + m_{\mathrm{vib}}^2 + \Sigma_{\text{eff}}(p_E^2; \langle \sigma \rangle, \langle \Sigma \rangle ) \right],
 \end{align}
where $\Sigma_{\text{eff}}$ denotes the effective self-energy, and the minus sign indicates a fermion loop.

The trace sums over the following modes:
\begin{enumerate}[label=\arabic*), leftmargin=2em]
\item The sum over Matsubara frequencies, with $p^0=i\omega_n+\mu_B/3$ for fermions, where $\omega_n=(2n+1)\pi/\beta$, $n\in\mathbb Z$, and $\beta=1/T$. For bosons, $\omega_n=2n\pi/\beta$.
\item Three-momentum modes, $\int \frac{d^3 p}{(2\pi)^3}$;
\item Vibration modes, $m_{\mathrm{vib}}^2$;
\item Spin states, $2s +1$.
\end{enumerate}
Detailed numerical analysis is necessary to obtain a phase diagram and identify the phase boundaries. For this purpose,
the method developed by Kenichi Aoki and collaborators~\cite{Aoki} seems very useful. They reformulate the functional renormalization group equations of Refs.~\cite{FRG1,FRG2} in terms of ordinary differential equations for the running couplings and fields. This allows us to find even the singular behavior of the fields (or condensates) as functions of the coupling constants.

Another potential advantage of dual QCD is that the gauge fields are integrated out, so the original gauge-field-dependent Dirac determinant does not explicitly appear. This reformulation may offer a different approach to the sign problem at finite baryon density, although the positivity of the resulting functional measure remains to be investigated.

\item Diquark condensation, such as 2SC (two-flavor color superconductivity) or CFL (color-flavor-locked phase)~\cite{CFL}, can be studied in our dual QCD with NJL-type interactions. Such condensation is modified by the dual gluon fluid in the strong-coupling expansion. In this setting, we can re-examine the neutron-star equation of state.
\end{enumerate}

\section{Prospects for future work}
Finally, we mention several questions that are difficult to answer at present but are expected to be resolved in the future.
\begin{enumerate}[label=(P\arabic*), ref=P\arabic*, leftmargin=2.6em]

\item A four-Fermi interaction is expected to become renormalizable when it couples appropriately to hydrodynamics. This is because dual QCD is obtained simply by Gaussian integration of the renormalizable theory of QCD.
\item A method for assigning quantum numbers to $p$-branes ($p > 1$) is needed (see the difficulties discussed for membranes in Ref.~\cite{Membrane}). In dual QCD, the gauge principle can be expressed through the \emph{Eulerian view} of hydrodynamics.

\item \emph{Nambu's problem}. He expressed the 2-form stream function in terms of two three-dimensional vectors:
\[
W_{0i}^a=-e_i^a,\qquad W_{ij}^a=\epsilon_{ijk}b^{ak},
\]
\[
V_0^a=\nabla\cdot\mathbf b^a,\qquad
\mathbf V^a=\dot{\mathbf b}^a+\nabla\times\mathbf e^a.
\]
Then, $\mathbf e^a$ is non-dynamical, while $\mathbf b^a$ is dynamical and propagates as a massless spin-1 particle with longitudinal modes.\footnote{
In the standard massless Kalb--Ramond theory
in four dimensions, only one physical degree
of freedom propagates, which is dual to a
massless scalar field. This degree of freedom
can be represented by a longitudinal mode
of the spatial vector $\mathbf{b}^a$.
The physical particle interpretation in the
present non-Abelian dual QCD model remains
to be clarified.
}

In a discussion with one of the authors (A.S.), Nambu posed the challenge of identifying the particle $\mathbf b$ that guarantees the existence of the Kalb--Ramond symmetry.
One possible answer may be found in a densely condensed vortex state. Detailed observations of neutron stars are therefore anticipated. Another possibility is the high-energy limit, where a perfect fluid appears. We therefore expect heavy-ion collision experiments to provide clues to Nambu's question.

\item Using the Stueckelberg trick, we may convert all current--fluid interactions into current--current interactions. All interactions are then of the current--current type. This suggests the possibility of studying QCD through a Sugawara-type construction~\cite{sugawara}, based on current algebras and the Lorentz algebra with Schwinger terms.
\end{enumerate}

\section*{Acknowledgements}
We are grateful to Shiro Komata for reading the manuscript and providing valuable comments. We thank the members of the OUJ Field Theory Collaboration, especially Ken Yokoyama for his understanding and interpretation of this paper.

\vspace{0.5\baselineskip}

\end{document}